%% file: paper.tex
\pdfoutput=1 
\documentclass{JINST}
\usepackage{graphicx} 
\usepackage{dcolumn}  
\usepackage{longtable}
\usepackage{bm}       
\usepackage{subfigure}
\usepackage{multirow} 
\title{Efficient, reliable and fast high-level triggering using a bonsai boosted decision tree.}
\author{V. V.\ Gligorov$^1$ and M. Williams$^{2,a}$\\ 
$^1$Organisation Europ\'{e}enne pour la Recherche Nucl\'{e}aire, Geneva, Switzerland \\
$^2$Imperial College London, London SW7 2AZ, UK\\
$^a$Current address: Massachusetts Institute of Technology, Cambridge, MA, USA
}

\abstract{ 
High-level triggering is a vital component in many modern particle physics experiments.  This paper describes a modification to the standard boosted decision tree (BDT) classifier,
the so-called {\em bonsai} BDT, that has the following important properties: it is more efficient than traditional cut-based approaches; it is robust against detector instabilities,
and it is very fast. Thus, it is fit-for-purpose for the online running conditions faced by any large-scale data acquisition system.
}
\begin{document}

\input{intro.tex}
\input{method.tex}
\input{toyex.tex}
\input{discussion.tex}


\section{Summary}

The BBDT algorithm presented in this paper has the following important properties: it is more efficient than traditional cut-based approaches; it is robust against detector instabilities, and it is very fast. Thus, it is fit-for-purpose for the online running conditions faced by any large-scale data acquisition system.  The BBDT has been proven to work at the LHCb experiment at CERN.  It should be considered for use in any current or future HLT algorithm.

\acknowledgments
We would like to thank our colleagues at the LHC$b$ experiment, especially Roel Aaij, Johannes Albrecht, Vanya Belyaev, Hans Dijstra, Ulrik Egede, Richard Jacobson and Gerhard Raven.
MW was supported in this work by STFC grant number ST/H000992/1. VVG was supported in this work by a Marie Curie Action: ``Cofunding of the CERN Fellowship Programme (COFUND-CERN)'' of the European Community's Seventh Framework Programme under contract number (PCOFUND-GA-2008-229600).

\input{bib.tex}
\end{document}

%% file: intro.tex
\section{Introduction}

The increase in data-production rates in particle physics experiments has led to an increase in the importance of high
level trigger (HLT) algorithms performing partial or even full event reconstruction.
This increase in data production has been caused by the high collision rates and/or
luminosities provided by modern accelerators and the large event sizes provided by modern detectors.
The resources required to store and process these large volumes of data do not exist; thus, trigger systems are
employed to reduce the data to manageable levels {\em online}, {\em i.e.}, during the running of the experiment.
This reduction is achieved by preferentially selecting certain events thought to be of particular interest
for further analysis, which we shall refer to as ``signal'' from here on.

First level (L0) triggers, {\em i.e.}, triggers implemented purely as part of a detector's hardware,
have been used since the earliest days of particle physics experiments. In their simplest form, as used in early bubble chambers,
L0 triggers make their decision based on the accelerator clock and do not discriminate between different event types.
In higher data rate environments they also use information from the detector itself in order to preferentially select
events containing signal.  An early example of this is the Cronin and Fitch CP-violation experiment which
triggered on the coincidence of signals in the scintillator and Cerenkov subdetectors~\cite{ref:l0}.
The advantage of L0 triggers is that they are fast and that their efficiency on signal events can be reliably modeled. 
The disadvantage is that they can only use low-level information; thus, they are only efficient
if (a) the signal has a very clear discriminating trait from the background and (b) that trait is accessible by
reading out a small number of electronic signals from the detector (without any significant algorithmic processing).
Unfortunately, conditions (a) and (b) are often times not satisfied.  In such situations HLT algorithms must be used.


The distinction between L0 and HLT algorithms is that the former makes decisions based on event information from a single subdetector or
simple coincidences between subdetectors, while the latter performs an online event reconstruction and makes a decision based on the full
event information.  Due to timing constraints,  the online reconstruction often differs from the offline one, {\em e.g.}, using a simplified detector
geometry or a coarser granularity.
An early example of an HLT based trigger was that of the ACCMOR collaboration~\cite{Daum:1981tw} which
used Cerenkov information to separate kaons from pions and momentum information to select events containing $\phi$ mesons based
on the invariant mass of the $K^+K^-$ pairs. This trigger is also notable as it was implemented using commercially-available
network-linked PCs. 

To-date, HLT algorithms have been cut-based and designed to
accomodate the inevitable detector instabilities that occur during online running. In this context, cut-based means that
while the HLT has access to the full detector information, correlations between the different discriminating variables are largely
ignored and an event is selected based on a series of individual criteria combined through a chain of logical ANDs.
So the HLT might look for a signal with high transverse momentum AND high displacement from a
primary interaction AND high invariant mass. 
There are, however, experiments currently running, and many more planned for the near
future, where cut-based HLT algorithms simply are not powerful enough to discriminate between signal and background.
Multivariate classifiers, {\em e.g.}, neural networks and boosted decision trees (BDTs), are known to be more powerful than
cut-based selections. There are, however, some very important concerns regarding using a BDT in an HLT algorithm (these are discussed in
detail in Sec.~2). 

In this paper we introduce a modification to the standard BDT algorithm that, by construction, addresses
these concerns making this algorithm fit-for-purpose for use in a HLT.
This HLT algorithm was designed for the LHCb experiment at CERN~\cite{ref:lhcbtrig}.  In 2011 the proton-proton collision rate at the
LHC was more than 10~MHz.  With event read-out sizes of $O(100~{\rm kB})$, this means that the LHCb detector was
exposed to $O(1~{\rm TB})$ of data per second.  Clearly, the data must be reduced online before it is put into permanent storage
or used for analysis. Many of the signals of interest cannot be separated from the massive LHC backgrounds using only
information available at L0.  Furthermore, these signals cannot be separated from the backgrounds using single-track algorithms
or even by cut-based multi-track ones. Thus, a multivariate classifier must be used.

The BDT HLT algorithm presented in this paper has been running since the start of LHCb data taking in 2011. 
Its performance has been excellent. We will refer to the LHCb HLT algorithm as an example; however, the algorithm itself is not LHCb specific.
This paper has been written to illustrate how this algorithm works so that it may be used by other experiments.
Section~2 describes the basic idea behind the algorithm.  Section~3 provides a toy-model example to illustrate how to use the
algorithm and also benchmarks its performance against a cut-based approach.  In Sec.~4 we provide some discussion on the
algorithms performance at LHCb  
before summarizing in Sec.~5.

%% file: method.tex
\section{The Bonsai Algorithm}

Decision trees (DTs) are multivariate classifiers that are built by looping over the variates and repeatedly performing one-dimensional splits of the data.
The criteria that determines where to split the data involves maximizing a chosen figure of merit (FOM).
For example, a commonly used FOM in particle physics is the so-called {\em signal significance} $S/\sqrt{S+B}$, where $S$ and $B$ are the number of signal and background events, respectively.  

To limit the effects of {\em overtraining} (fine tuning of the DT structure that often occurs during training due to the finite size of the training data sample),
DTs are often {\em boosted}.  One example of a boosting algorithm is {\em bagging}~\cite{ref:bag}.  Bagging involves making a large number of bootstrap copy training samples
by sampling with replacement from the original.  One then trains an independent BDT on each bootstrap copy sample; the BDT response is then the fraction of
these DTs in which an event is in a signal {\em leaf} (as opposed to a background one).  This procedure greatly enhances the power of the DT.  We note here
that any FOM and any type of boosting can be used in conjunction with the algorithm described below; these choices were provided to help illustrate how BDT algorithms work.
For a good review of BDTs, see Ref.~\cite{ref:bdt}.  

Multivariate classifiers work by defining $n$-dimensional regions of the multivariate space as signal or keep regions by learning from the training data samples provided to them.
There are three major concerns which need to be addressed prior to using such a classifier in an HLT algorithm:
\begin{itemize}
\item If the keep regions are small relative to the resolution or stability of the detector, the signal could oscillate in and out of the keep regions.
This would result in, at best, a less efficient trigger and, at worst, a trigger whose efficiency is very difficult to understand.  
\item In many cases the signal samples by necessity must come from simulations because the signals have, in fact, not yet been observed in data.
In other cases the trigger is meant to be {\em inclusive}, {\em i.e.}, the trigger is meant to select classes of signal types rather than one specific signal channel. 
In both cases, the signal PDFs might not be completely accurate or even available during the training process. 
\item Any HLT algorithm must run in the online environment; thus, it must be extremely fast.    
\end{itemize}
The simplest way to address these concerns is to discretize all of the variables used in the BDT.
This limits where the splits of the data can be made and, in effect, permits the grower of the tree to control and shape its growth; thus, we are calling this a bonsai BDT (BBDT).

This technique works by enforcing that the smallest keep interval that can be created when training the BBDT is
\begin{equation}
  \Delta x_{\rm min} > \delta_x ~\forall ~x~ {\rm on~all~leaves},
\end{equation}
where,
\begin{equation}
 \delta_x = {\rm MIN} \{ |x_i - x_j| : x_i,x_j \in x_{\rm discrete} \}.
\end{equation}
The constraints that govern the choice of $\{ x_{\rm discrete} \}$ and ensure that the concerns of using a BDT in an HLT algorithm are then as follows:
\begin{itemize}
\item $\delta_x$ should be greater than the resolution on $x$ in the detector and should be large with respect to the expected online variations in $x$.  
\item The discretization should reflect what properties of the training signal PDFs that are to be incorporated in the BBDT. For example, if the BBDT is meant to be inclusive,
then the discretization should help ensure that the BBDT learns the common traits shared by all signals of a given type rather than learning
a large sum of traits specific to the signal samples used in the training.
\item Discretization means that the data can be thought of as being binned, even though many of the possible bins may not form leaves in the BBDT;
thus, there are a finite number, $n_{\rm keep}^{\rm max}$, of possible keep regions that can be defined.  If the $n_{\rm keep}^{\rm max}$ BBDT response
values can be stored in memory\footnote{If there is not enough memory available to store all of the response values, there are a number of simple alternatives that can be used.  For example, if the cut value is known then just the list of indices for keep regions could be stored.}, 
then the extremely large number of {\rm if/else} statements that make up a BDT can be converted into a one-dimensional array
of response values.  One-dimensional array look-up speeds are extremely fast; they take, in essense, zero time.
\end{itemize}
Thus, by construction the BBDT is fit for purpose for use in an HLT algorithm and addresses all of the concerns listed above.

%% file: toyex.tex
\section{Toy Model Example}

Having described the method, we will now demonstrate its superiority over a cut-based selection algorithm
using a simplified situation in which only two discriminants are available : the transverse momentum\footnote{Transverse here means perpendicular to the beam axis.} ($p_T$)
of the candidates, and their displacement from the primary interaction. Although simplified, the signals
and backgrounds used reflect, as much as possible, the actual situation faced by the LHCb experiment
in triggering on generic decays of $B$ mesons.

\subsection{Trigger discriminating variables}
The simulated trigger models a situation in which a proton-proton collision has been reconstructed in the tracking system
of a detector. Only charged tracks are considered to form part of the event. The $p_T$ of each track
is assumed to be reconstructed with no uncertainty\footnote{LHC tracking systems universally achieve sub percent
precision for this quantity at the HLT reconstruction stage.}. The perpendicular distance of
closest approach to the primary interaction for each track, henceforth impact parameter (IP), is calculated from the event geometry.
The IP of each track is measured with a precision that
varies with the track $p_T$; larger $p_T$ tracks are more precisely reconstructed. The resolution
is implemented as a Gaussian smearing of the actual track IP.

The simulated trigger makes its decisions based on forming a three track vertex. All possible combinations of tracks are assumed
to be available to the trigger, and at least one combination must satisfy the trigger algorithm in order for the event to pass.
The information available to the trigger is the sums of the $p_T$ and IP of the tracks making up the vertex.

\subsection{Toy Model Data: Signal}
The signal events are $B$ meson decays. Two different decays are modelled: a decay into four charged tracks (two kaons and two pions),
and a decay into five charged tracks (two kaons and three pions). The momentum of the $B$ mesons is sampled from PYTHIA~\cite{ref:pythia},
while their lifetime is taken to be $1.5$~ps. The $B$ mesons are decayed using the {\tt TGenPhaseSpace} class of
the ROOT~\cite{ref:ROOT} software framework and assumed to contain no intermediate resonances. 

\begin{figure}[tbh]
 \centering
 \includegraphics[width=.99\textwidth]{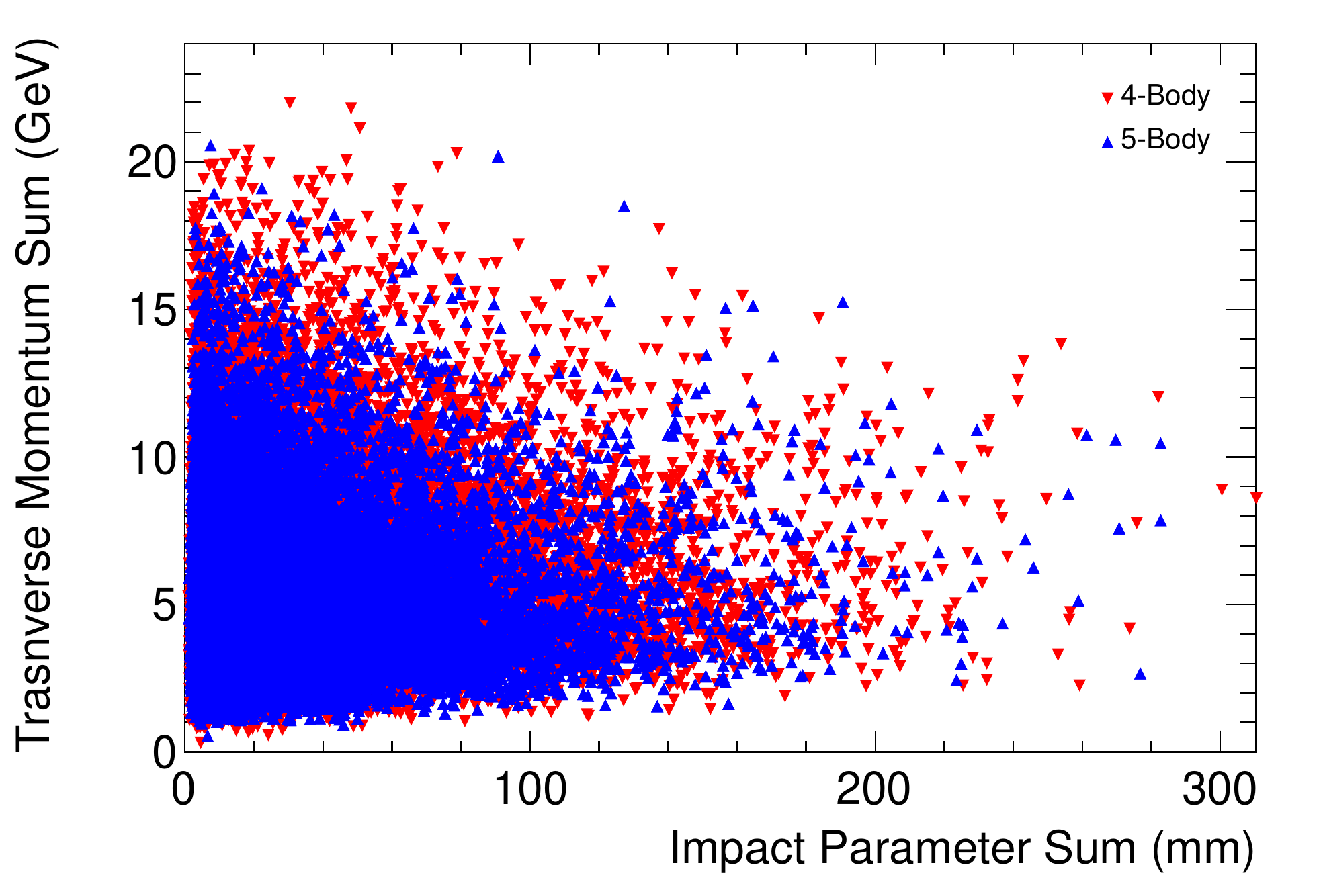} 
 \includegraphics[width=.99\textwidth]{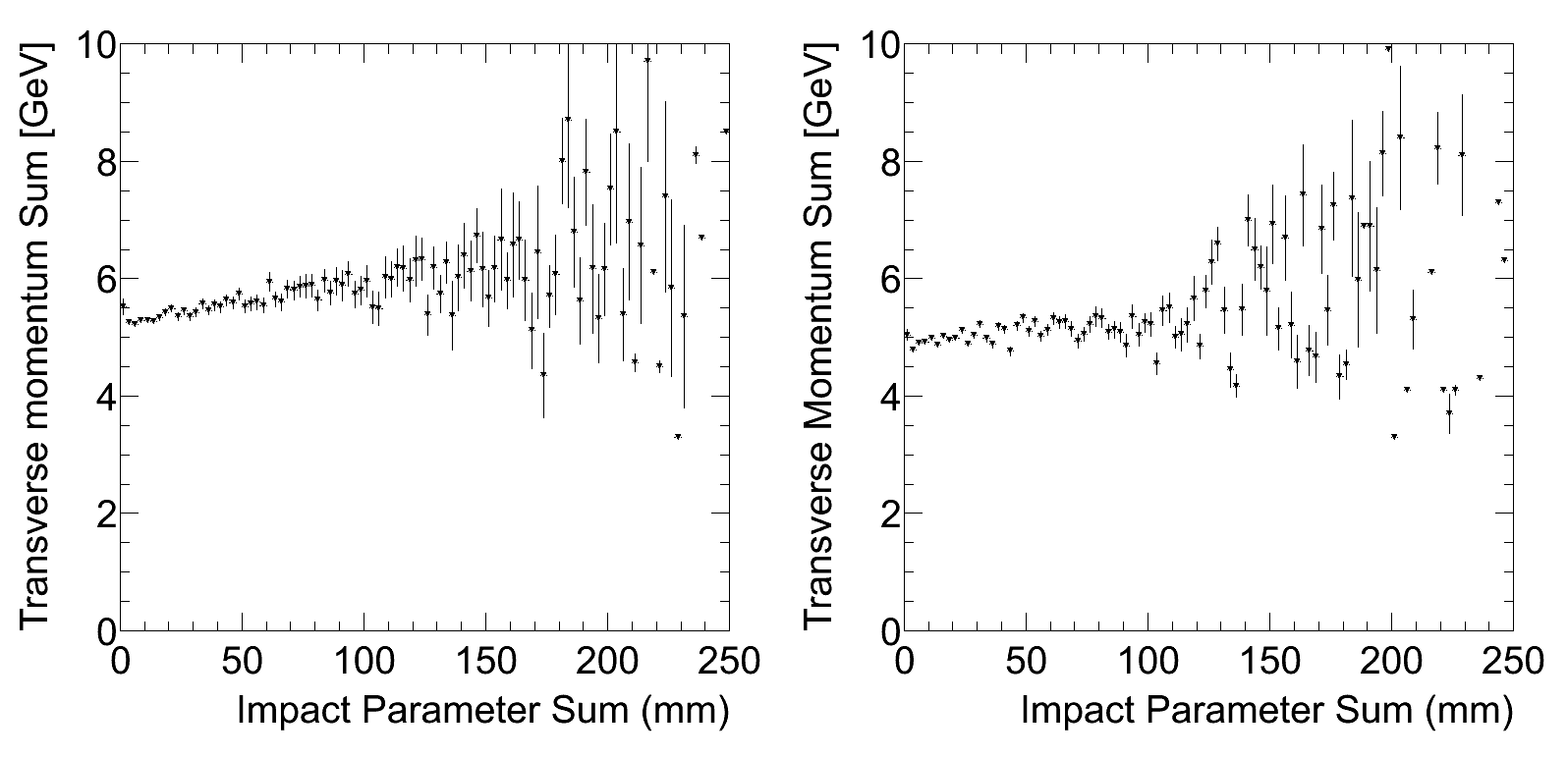}
 \caption{Top : the discriminating variable distributions for the signals. Bottom : the average transverse momentum sum as a function of the IP sum for the four- and five-body signals (left and right respectively).}
 \label{fig:SignalDists}
\end{figure}

Two different signals are generated to test the inclusive nature of the algorithm; the algorithm will be trained on the four-body
decay and tested on both the four- and five-body decays. 
In order to make the five-body signal a tougher challenge, its lowest transverse momentum track is discarded prior to forming the three-track
vertex candidates for the trigger to consider. This reflects the reality that low transverse momentum tracks suffer greater multiple scattering and are
harder to reconstruct, especially in the time-limited online environment. The result is that some candidates that could have otherwise fired the trigger, {\em e.g.}, because of the exceptionally large IP of this low momentum track, are removed from consideration.
The signal distributions are shown in Fig.~\ref{fig:SignalDists}.  The five-body signal candidate vertices generally have smaller $p_T$ and IP than the four-body ones.

\subsection{Toy Model Data: Backgrounds}

Three types of backgrounds are simulated: pure combinatorial background, consisting of three random tracks originating directly from the primary interaction
and, therefore, having a true IP of zero; so-called {\em ghost} background consisting of two tracks from the primary interaction plus one fake
track or {\em ghost}; and prompt charm backgrounds, consisting of two tracks from the decay of a $D$ meson combined with a fake track. These backgrounds reflect the reality
of the LHC environment in which one is fighting against multiple fake signatures each of which has its own distribution in the discriminating variables.

The inclusion of fake tracks is important because any tracking algorithm produces some proportion of ghosts which are empirically observed to have an almost
random measured IP.   The measured IP of ghost tracks is modeled here with a very long-lived\footnote{The half-life is
set to 250~mm to be compared to the distributions in Fig.~\ref{fig:SignalDists}} exponential. These ghosts have a similar $p_T$ distribution to genuine
tracks, and present a very challenging background, especially when combined with the decay products of genuinely long-lived $D$ mesons. The latter are modeled
in an analogous way to $B$ mesons; the $D^+$ is chosen as it is the most challenging to reject, with a lifetime of $\sim 1$~ps.

\begin{figure}[tbh]
 \centering
 \includegraphics[width=.99\textwidth]{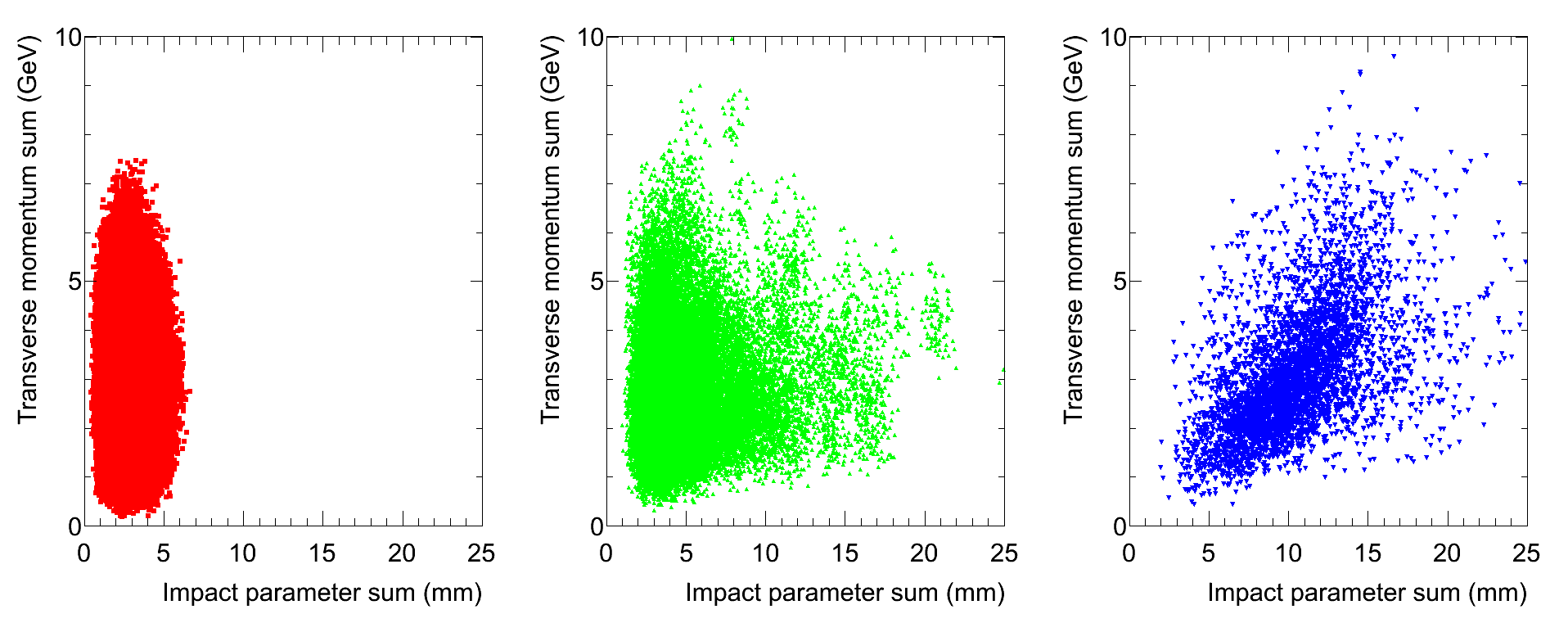}
 \caption{Discriminating variable distributions for the backgrounds. From left to right : pure combinatorial, ghost background, prompt charm.}
 \label{fig:BackgroundDists}
\end{figure}

Every background event is simulated with 30 tracks originating from the primary interaction, of which $10\%$ are ghosts, as well as a single $D^+\to K^- \pi^+ \pi^+$ decay.
The background distributions are shown in Fig.~\ref{fig:BackgroundDists}. Note that there is an enormous difference in the relative abundances of these three background types,
with the pure combinatorics being by far the most abundant and the ghost plus charm background the least abundant (but most dangerous).

\subsection{Performance}

For each of the HLT algorithms defined below, the optimal selections are determined using {\em training} samples of the 4-body signal and background.  Their performances are then evaluated using {\em validation} samples of the 4-body signal, 5-body signal and background.  The optimal selections are simply those that maximize the efficiency on the 4-body signal while achieving a factor of 100 reduction in the background.  
The stability of each HLT algorithm is tested using validation samples that include an additional $1\sigma$ smearing of the IP.  

Three HLT algorithms are studied: cut-based, BDT-based and BBDT-based.  The performance of each is given in Table~\ref{tab:results}.  As expected, the BDT and BBDT are much more efficient than the cuts.  This simple example only has two variables, but the relative efficiency of the (B)BDT-based HLT is  (19\%)22\% higher than the cut-based one.  
The BBDT is 3\% less efficient than the BDT on the 4-body signal; however, it is 1\% more efficient on the 5-body signal.  Whatever the BDT has learned about the 4-body signal to give it the slight advantage over the BBDT must not be a common trait shared by the the two types of signal.  

\begin{table}
\caption{\label{tab:results} Performance on the toy model data of a cut-based, BDT-based and BBDT-based HLT algorithm. $\epsilon_{n-{\rm body}}$ are the efficiencies on the $n$-body signals. The instability is the increase in the rate under the imperfect online conditions (see text for details).}
\begin{center}
\begin{tabular}{c|cc|c}
  type & $\epsilon_{4-{\rm body}}$ & $\epsilon_{5-{\rm body}}$ & instability \\
  \hline
  cuts & 63\% & 55\% & 9\% \\
  BDT & 77\% & 68\% & 55\% \\
  BBDT & 74\% & 69\% & 10\% 
\end{tabular}
\end{center}
\end{table}

The most important column in Table~\ref{tab:results} is the one showing the instability of the rate under imperfect online running conditions.  The BDT rate increases 6 times more than the cut-based rate does; however, the BBDT rate is very close to the cut-based one.  For an experiment with this level of possible instability online, the BDT performance here is unacceptable.  Of course, if one knew how much instability to expect the training data could just be smeared to account for it.  Unfortunately, online instabilities often fall into the category of Rumsfeldian uncertainties.  They are expected to occur but where and by how much is unknown; thus,  it is very difficult to produce training samples that guard against all possible online instabilities.  
This section demonstrates that the BBDT is almost as stable as cuts while being almost as efficient as a BDT.   This makes it ideal for an HLT algorithm.  

This BBDT had 10 allowed split points per variable.  The smallest allowed keep region that could be defined was $5\sigma$ wide (in the lowest region of IP).  In general, more allowed split points gives higher efficiency but also introduces more sensitivity to online instabilities.  When defining the allowed split points, the grower of the BBDT has total control over the size of the smallest possible keep regions that can be defined.  Typically, it is possible to define the split points such that online instabilities will not be an issue without the need for complicated simulation studies.    

%% file: discussion.tex
\section{Performance at LHCb}

The performance of the BBDT algorithm as actually deployed in the LHCb experiment is described in detail in~\cite{ref:lhcbtrig}.
Without duplicating any results, we would emphasize the following points. The BBDT is implemented for the so-called {\em topological} trigger
that aims to select any $B$ meson decay which produces charged tracks (kaons, pions, protons, muons or electrons). It does so
by reconstructing two-, three-, and four-track vertices displaced from the primary interaction and then calculating the BBDT response
based on the properties of this vertex and the properties of the tracks from which the vertex has been reconstructed. 

Since its deployment,
comprising about 98$\%$ of the luminosity collected to-date, the BBDT topological trigger has served as the main trigger for $B$ physics on LHCb.
During this time it has undergone no fundamental changes or retunings. In particular, it has proven robust against the misalignments present at the
start of each calendar year (introduced following hardware interventions during the winter shutdowns of the accelerator), and against the change from $7$ to $8$~TeV centre
of mass energy of the collisions. 

The BBDT topological trigger has been measured in a data-driven manner to be highly efficient on a variety of $B$ decay topologies.  Its efficiency
has been measured as a function of various $B$ meson properties  and has displayed no pathological structures indicative of overtraining or resolution effects.
Futhermore, as described in the LHCb upgrade letter of intent~\cite{ref:LOI}, the BBDT topological trigger has been verified in simulation
to maintain its performance at $14$~TeV centre of mass energy and a $25\%$ greater per-bunch luminosity.   
We can therefore conclude that the BBDT achieves in practice its stated aims of being efficient on signal, powerful against background, robust
against mis-reconstruction, and understandable in its behavior.